# FINAL COOLING WITH THICK WEDGES FOR A MUON COLLIDER*

D. Fu†, A. Badea. K. Folan Di Petrillo, University of Chicago, Chicago, IL USA
D. Neuffer, D. Stratakis‡, Fermilab, Batavia, IL USA

*Abstract*

In the final cooling stages for a muon collider, the transverse emittances are reduced while the longitudinal emittance is allowed to increase. In previous studies, Final cooling used absorbers within very high field solenoids to cool low-momentum muons. Simulations of the systems did not reach the desired cooling design goals. In this study, we develop and optimize a different conceptual design for the final 4D cooling channel, which is based on using dense wedge absorbers. We used G4Beamline to simulate the channel and Python to generate and analyze particle distributions. We optimized the design parameters of the cooling channel and produced conceptual designs (corresponding to possible starting points for the input beam) which achieve transverse cooling in both x and y by a factor of ≈ 3.5. These channels achieve a lower transverse and longitudinal emittance than the best previously published design.

## INTRODUCTION

The Snowmass evaluation of future high-energy physics needs followed by the P5 report asserted the long term priority of a 10 TeV scale collider such as a 10 TeV muon collider. A muon collider is a particle accelerator that generates collisions of $\mu^+$ bunches with $\mu^-$ bunches. Figure 1 shows a possible layout of a high-energy muon collider facility as it might be developed on the Fermilab site [1]. It includes a proton source that obtains high-intensity proton bunches that collide onto a target, producing pions that decay to muons, which are collected and cooled. The resulting muon bunches are then accelerated and transferred to a fixed energy collider ring for multiturn collisions.

Obtaining high luminosity requires cooling muons to small emittances in a multistep cooling channel. Figure 2 shows the progress in emittance reduction thorugh the cooling channel through the Final Cooling (circled in blue). In previous studies, Final Cooling used absorbers within very high field solenoids to cool low-momentum muons [2]. The baseline transverse emittance goal in Final Cooling is ≈ $25\mu$; the previous studies obtained ≈ $50\mu$ .

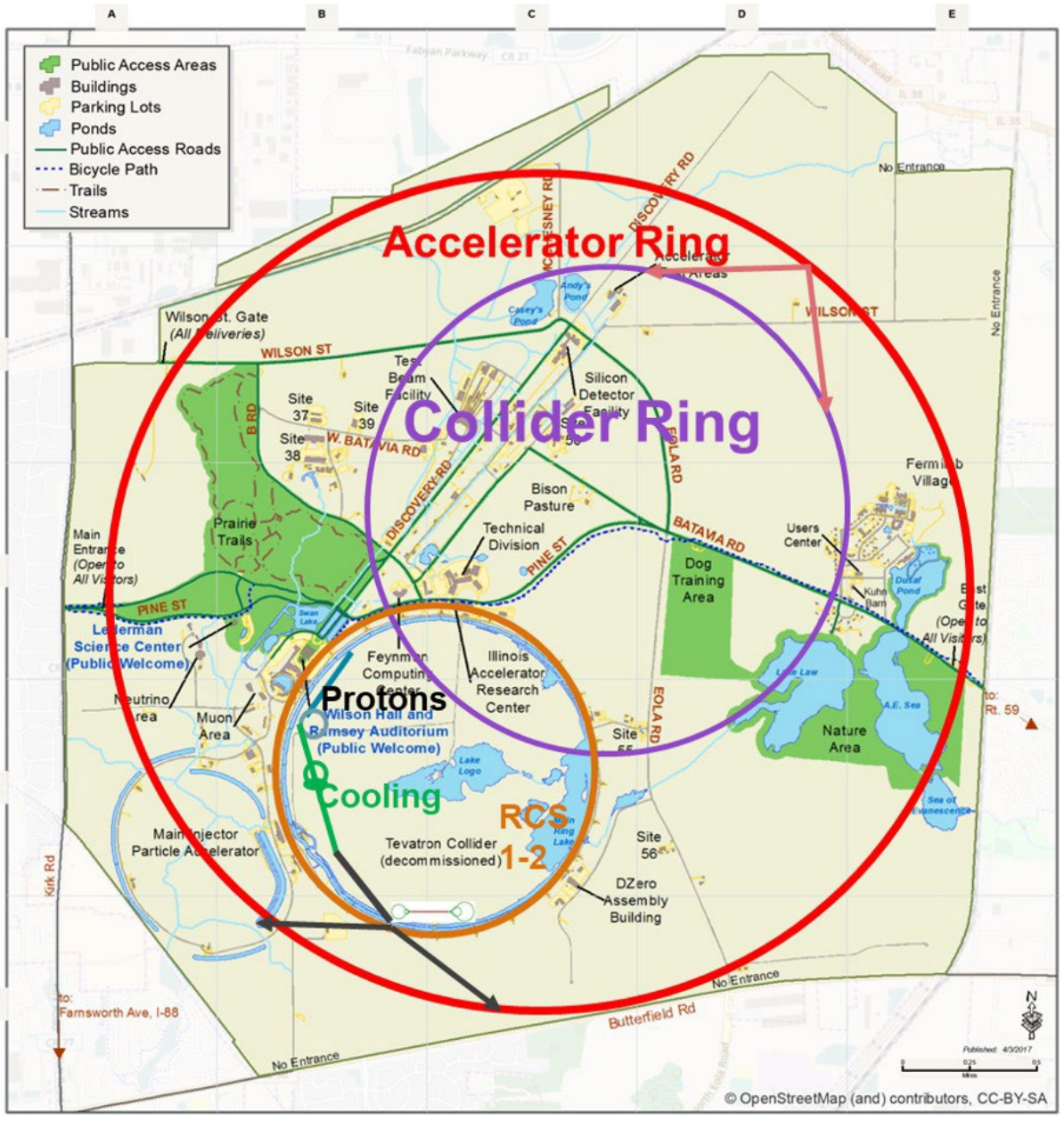

Figure 1: Diagram of how a muon collider facility could be built on the Fermilab site. The facility is designed for 10 TeV collisions (5 TeV $\mu^+$ and $\mu^-$).

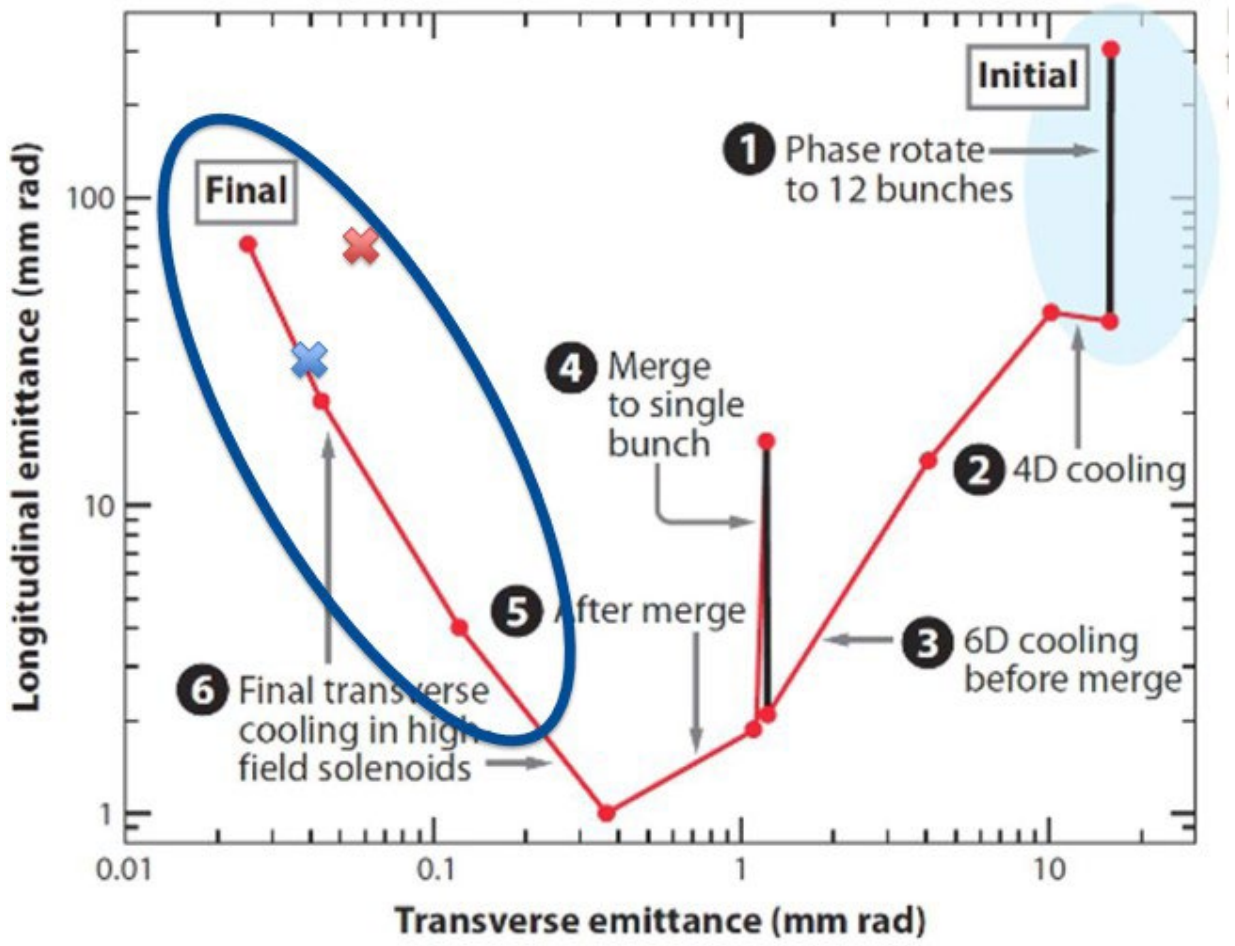

Figure 2: Path of the muon cooling process in transverse and longitudinal emittances. The Final Cooling step we investigated is circled in blue. The previous best result is marked in red, and our best result is marked in blue.

## WEDGE-BASED FINAL COOLING SCENARIO AND STUDY

In this study, we develop and optimize a different conceptual design for the final cooling channel, which is based on single thick wedges [3]. This method passes the beam through a wedge of solid material, which absorbs energy from the muons through ionization energy losses. Because of the linear variation in thickness, the wedge introduces linear dispersion, which is the variation of the position of the particles with the momentum of those particles. This dispersion is then removed using bending magnets, which displaces off-momentum particles towards the centerline of the beam, reducing the beam size, and correspondingly the transverse emittance. This process also increases the spread in momenta of the muons (measured by the standard deviation of momentum, $\sigma_p$). The beam is then sent through a drift plus rf cavity region for energy-phase rotation, which decreases $\sigma_p$ at the cost of increasing $\sigma_t$ (the spread of the particles in time). The beam is then passed through a second wedge to reduce the emittance in the other transverse axis. The layout of the cooling channel is shown in Fig. 3.

* Work supported by Fermi Research Alliance, LLC under Contract No. DE-AC02-07CH11359 with the U.S. Department of Energy, Office of High Energy Physics.
† Danielfu@uchicago.edu
‡ diktys@fnal.gov

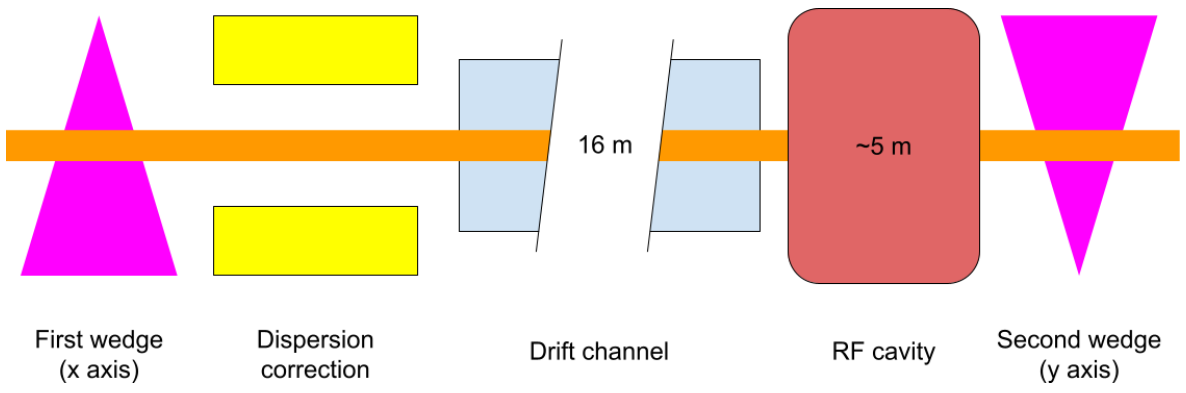

Figure 3: Diagram of the wedge cooling pathway investigated.

We considered two starting points for our final cooling channel: 145 µm transverse emittance with 1.0 MeV/c $\sigma_p$, and 110 µm transverse emittance with 0.8 MeV/c $\sigma_p$. These correspond to the outputs of the B10 and B11 stages in a previously designed ionization cooling channel using high-field solenoids [4]. Our channel could therefore replace the final stages of this design, further reducing the transverse emittance while avoiding the need for extremely high-field solenoids.

We wrote a Python script to generate particle distributions with specified mean momentum ($p$), emittances ($\epsilon_x$, $\epsilon_y$, $\epsilon_z$), Twiss parameters ($\beta$, $\alpha$, $\gamma$) in the transverse axes, and standard deviation of momentum ($\sigma_p$), following an algorithm described by Mike Syphers [5]. We used G4Beamline [6], a Geant4-based simulation software, to model the effects of the wedges and RF cavity on these distributions. The geometry of the wedges is defined by two parameters: the wedge length (measured at the point where the centerline of the beam crosses the wedge) and the half-angle. Figure 4 shows how these parameters are defined.

## OPTIMIZED RESULTS

We optimized the design parameters of the cooling channel and produced two conceptual designs (corresponding to the two possible starting points for the input beam) which achieve transverse cooling in both x and y by a factor of ≈ 3.5. These channels achieve a smaller transverse and longitudinal emittance than previously published designs.

### *Optimization Parameters*

A large number of potential parameter choices can be considered in developing optimized solutions and we considered several variations [7].

In both cases we initiated cooling at $p_\mu$ =100 MeV/c as a baseline starting point, where a large exchange can be obtained with a single dense wedge. That choice can be varied in future optimizations. For both wedges the length and angle of the wedge was varied over a broad range. The wedge material chosen was diamond (dense carbon), a dense low-Z material. (A low-Z material is needed to minimize multiple scattering ) Other materials can be considered. For the rf match between wedges the goal was to reduce the momentum spread t0 ≈ 1 MeV/c while lengthening the bunch, using an initial 16 m drift followed by 25 MHz rf at 5.1 MV/m gradient.

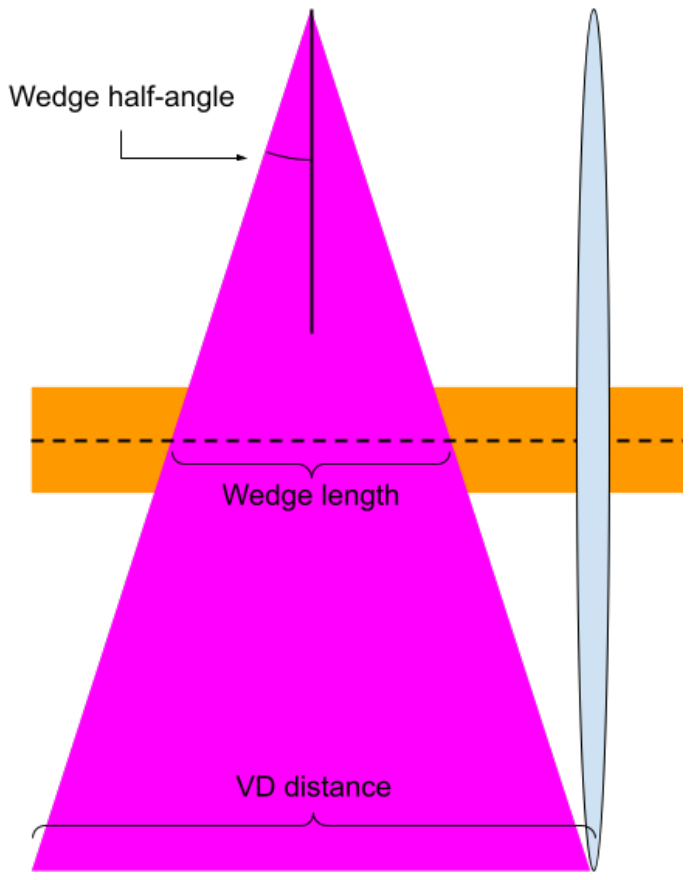

Figure 4: Diagram of the wedge geometry, with wedge length, half-angle, and virtual detector, where emittances are calculated. Note that wedge length is defined as the distance that the beam centerline intersects the wedge.

Table 1: Optimal Wedge Lengths, Wedge Angles, RF Lengths, and RF Gradients for Both Cases Considered

| Case | 145 µm | 110 µm |
|---|---|---|
| Wedge 1 length (mm) | 7.5 | 6.6 |
| Wedge 1 angle (deg) | 46.5 | 51.6 |
| Drift length (m) | 16.0 | 16.0 |
| RF cavity length (m) | 5.6 | 5.7 |
| RF gradient (MV/m) | 5.1 | 5.1 |
| Wedge 2 length (mm) | 6.1 | 5.1 |
| Wedge 2 angle (deg) | 40.8 | 47.2 |

We produced an optimized channel for both the 145 µm and 110 µm cases. Table 1 gives the optimized parameters. Figures 5 and 6 show the evolution of the phase-space distri-bution as the particles travel through the channels. Tables 2 and 3 give the emittances and average momentum various points along these channels. Both cases had ≈ 84 % muon beam survival through the transport, including decay and aperture losses.

Table 2: Beam parameters through the optimal channel for the 145 µm case. This simulation was run with 200,000 particles.

| Stage | $\epsilon_x$ (µm) | $\epsilon_y$ (µm) | $\epsilon_z$ (mm) | $p$(MeV/c) |
|---|---|---|---|---|
| Initial | 144.9 | 145.1 | 1.262 | 100.0 |
| First wedge | 45.9 | 151.4 | 6.319 | 87.2 |
| After RF | 37.7 | 139.5 | 4.401 | 86.3 |
| 2nd wedge | 44.5 | 50.1 | 29.77 | 72.6 |
| After 4$\sigma$ cut | 40.1 | 43.9 | 28.65 | 72.7 |

Table 3: Beam parameters through the optimal channel for the 145 µm case. This simulation was run with 200,000 particles.

| Stage | $\epsilon_x$ (µm) | $\epsilon_y$ (µm) | $\epsilon_z$ (mm) | $p$(MeV/c) |
|---|---|---|---|---|
| Initial | 109.8 | 109.4 | 1.39 | 100.0 |
| First wedge | 34.5 | 114.1 | 7.115 | 88.9 |
| After RF | 27.9 | 105.4 | 4.886 | 87.9 |
| 2nd wedge | 32.4 | 39.4 | 24.54 | 77.0 |
| After 4$\sigma$ cut | 29.9 | 36.3 | 23.65 | 77.1 |

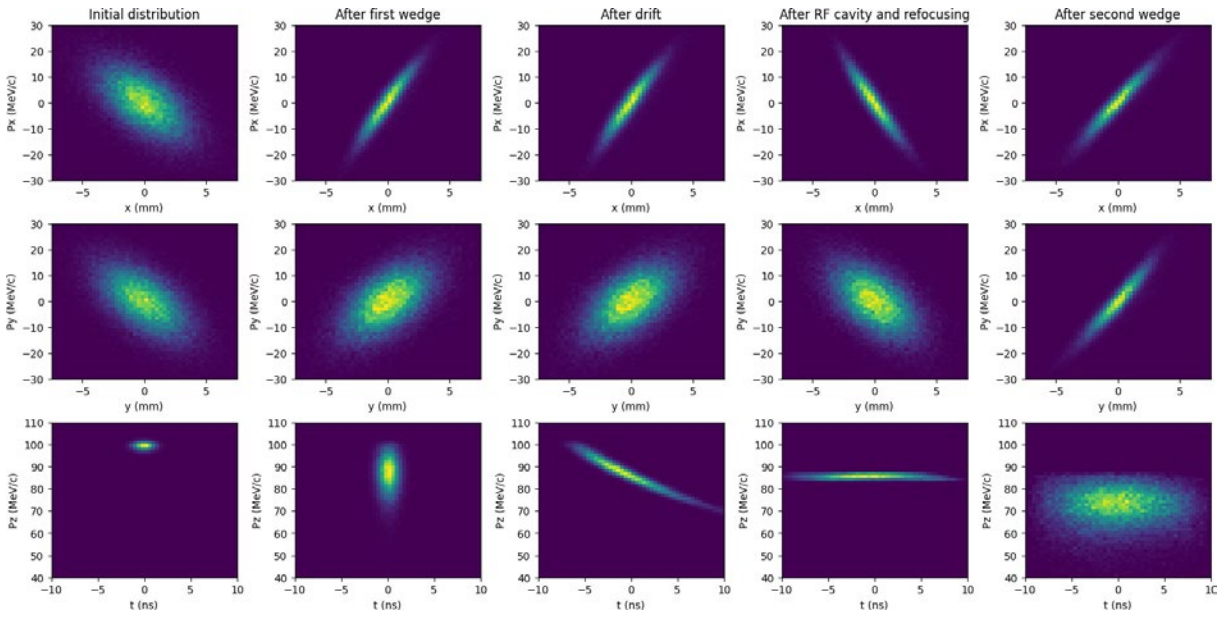

Figure 5: Phase-space distributions at various points along the optimized channel for the 145 µm case.

The first wedge in both cases achieves cooling by a factor of about 3.5 (145 µm to 40 µm and 110 µm to 32 µm). The second wedge performs less well than the first, only able to reduce $\epsilon_y$ to 43.5 µm in the 145 µm case and 33.8 µm in the 110 µm case. This is likely due to the higher $\sigma_p$ going into the second wedge, as the phase-rotation only reduces $\sigma_p$ to around 1.3 MeV/c (rather than the 1.0 MeV/c or 0.8 MeV/c that the beam began with). Improvements to the phase-rotation setup could therefore result in improved y-axis emittance reduction.

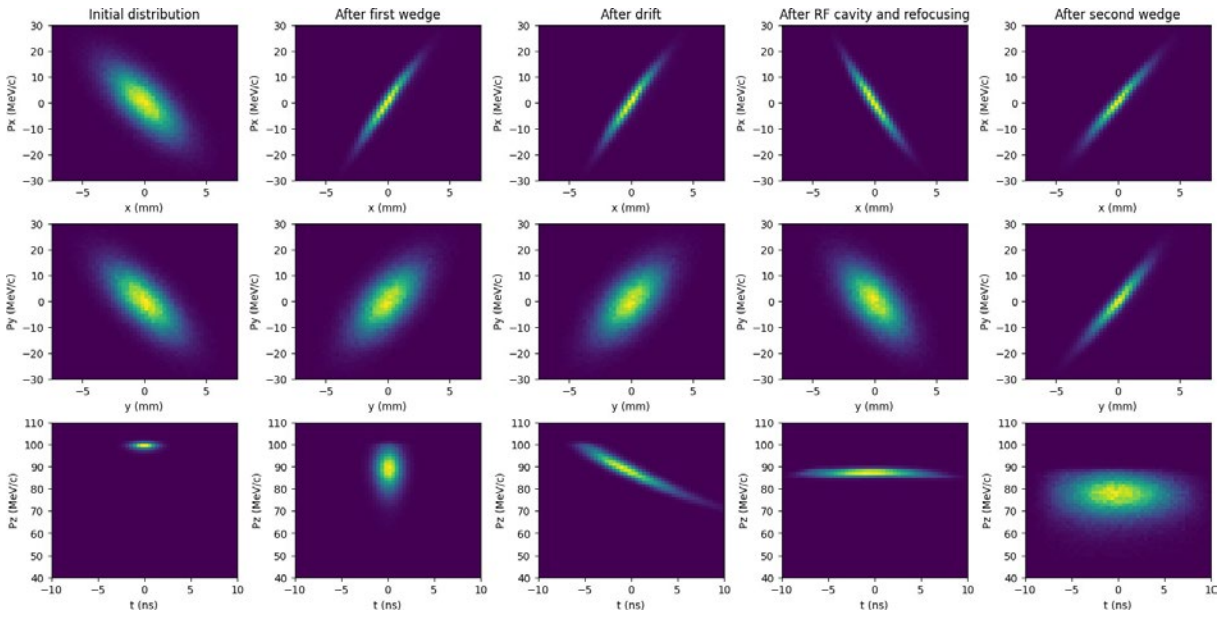

Figure 6: Phase-space distributions at various points along the optimized channel for the 110 µm case.

### *Future Required Development*

The present study will require further development before establishing a complete final cooling system. The initial conditions depended upon the output of the cooling channels of Ref. [4]. Other cooling channels, such as those of Ref. [2], have larger transverse or longitudinal emittances and would require somewhat different optimizations. The present example did not fully implement the transverse optics of the cooling channel. A full implementation would require specification of the transverse optical focusing elements which would include quadrupole and/or solenoidal magnets, dipoles and possibly correction sextupoles. The beams exiting the wedges achieve transverse cooling by transferring transverse positions into position plus dispersion times momentum offset. To obtain cooling the dispersion must be corrected by a combination of bending and focussing magnets (dispersion suppressor optics) in the drift length following the wedge. The next major effort will be in defining such optics solutions. The transverse optics must also include focussing through the rf systems and focussing to small beam sizes into the wedges with optical matching following the wedges.

Future designs should also consider integrating the wedge-based exchanges with other cooling approaches.

## CONCLUSION

We have developed parameters for wedge-based final cooling systems for a Muon Collider, based on sample final cooling designs. Directions for future development are discussed.

## REFERENCES

[1] P. V. Bhat *et al.*, “Future Collider Options for the US”, *arXiv*, 2022. doi:10.48550/arXiv.2203.08088

[2] H, Kamal Sayed, R. B. Palmer, and D. Neuffer, “High Field – Low Energy Muon Ionization Cooling Channel”, *Phys. Rev. ST Accel. Beams*, vol. 18, p. 091001, 2015. doi:10.1103/PhysRevSTAB.18.091001

[3] D. Neuffer, H. Kanmal Sayed, J. Acosta, T. Hart, and D. Summers, “Final cooling for a high-energy high-luminosity lepton collider”, *J. Inst.*, vol. 12, p. T07003, 2017. doi:10.1088/1748-0221/12/07/T07003

[4] D. Summers, “More Muon Cooling, Higher Luminosity”, 2020. https://indico.cern.ch/event/961803/contributions/4064621/attachments/2141709/3608895/DS_mu_12Nov.pdf

[5] M. Syphers, “Creation and Analysis of Beam Distributions”, 2017. https://nicadd.niu.edu/~syphers/tutorials/analyzeTrack.html

[6] T. J. Roberts *et al.*, G4beamline, http://g4beamline.muonsinc.com

[7] D. Fu, D. Neuffer, and D. Stratakis, “Final Cooling with Thick Wedges for a Muon Collider”, Fermi National Accelerator Laboratory (FNAL), Batavia, IL, United States, Rep. FERMILAB-FN-1236-AD-STUDENT, 2023. doi:10.2172/2246786